\documentclass[11pt]{JHEP3}

\JHEPspecialurl{http://jhep.sissa.it/JOURNAL/JHEP3.tar.gz}

\usepackage{epsfig,multicol}
\usepackage{graphicx}
\usepackage{amsmath,amssymb}
\usepackage{mathrsfs}

\newcommand\fverb{\setbox\fverbbox=\hbox\bgroup\verb}
\newcommand\fverbdo{\egroup\medskip\noindent%
			\fbox{\unhbox\fverbbox}\ }
\newcommand\fverbit{\egroup\item[\fbox{\unhbox\fverbbox}]}
\newbox\fverbbox

\newcommand{\pslash}{p\kern-1ex /}
\newcommand{\qslash}{q\kern-1ex /}
\newcommand{\lslash}{l\kern-1ex /}
\newcommand{\sslash}{s\kern-1ex /}
\newcommand{\kaslash}{k_a\kern-2ex /}
\newcommand{\kbslash}{k_b\kern-2ex /}
\newcommand{\Dslash}{\mathcal{D}\kern-1.5ex /}

\newcommand{\beqa}{\begin{eqnarray}}
\newcommand{\eeqa}{\end{eqnarray}}

\newcommand{\ba}{\begin{eqnarray}}
\newcommand{\ea}{\end{eqnarray}}
\newcommand{\be}{\begin{equation}}

\title{Effective potential from zero-momentum potential}

\author{\ \ \ \ J\'anos Balog$^{1,2}$\footnote{\tt balog.janos@wigner.mta.hu}
\ \ and\ \ 
Pengming Zhang$^1$\footnote{\tt zhpm@impcas.ac.cn} 
\\
\vskip 1ex
$^1$ {\it Institute of Modern Physics, 
Chinese Academy of Sciences, Lanzhou 730000, China}\\
\vskip 1ex
$^2${\it MTA Lend\"{u}let Holographic QFT Group, Wigner Research Centre} \\
{\it H-1525 Budapest 114, P.O.B. 49, Hungary}\\
}

\received{\today} 		
\accepted{\today}	

\abstract{We obtain the centre-of-mass frame effective potential from
the zero-momentum potential in Ruijsenaars-Schneider type 1-dimensional
relativistic mechanics using classical inverse scattering methods.
}

\begin{document}


\newcommand{\con}{\,\star\hspace{-3.7mm}\bigcirc\,}
\newcommand{\convu}{\,\star\hspace{-3.1mm}\bigcirc\,}
\newcommand{\Eps}{\Epsilon}
\newcommand{\gM}{\mathcal{M}}
\newcommand{\dD}{\mathcal{D}}
\newcommand{\gG}{\mathcal{G}}
\newcommand{\pa}{\partial}
\newcommand{\eps}{\epsilon}
\newcommand{\La}{\Lambda}
\newcommand{\De}{\Delta}
\newcommand{\nonu}{\nonumber}
\newcommand{\beq}{\begin{eqnarray}}
\newcommand{\eeq}{\end{eqnarray}}
\newcommand{\ka}{\kappa}
\newcommand{\ee}{\end{equation}}
\newcommand{\an}{\ensuremath{\alpha_0}}
\newcommand{\bn}{\ensuremath{\beta_0}}
\newcommand{\dn}{\ensuremath{\delta_0}}
\newcommand{\al}{\alpha}
\newcommand{\bm}{\begin{multline}}
\newcommand{\fm}{\end{multline}}
\newcommand{\de}{\delta}


\section{Introduction and motivation}

Recent advances in lattice QCD make it possible to measure relevant physical 
quantities at realistic, physical quark masses. This includes the measurement
of the nuclear force between nucleons by the HAL QCD collaboration
\cite{Ishii:2006ec,Aoki:2008hh,Aoki:2009ji}. The HAL QCD method is based on
the Nambu-Bethe-Salpeter (NBS) wave function defined by
\begin{equation}
\psi^{\rm NBS}_{\bf k}({\bf x})=\langle 0\vert N({\bf 0},0)N({\bf x},0)
\vert {\rm NN};{\bf k}\rangle^{\rm in}, 
\end{equation}
where $\langle 0\vert$ is the QCD vacuum state, 
$\vert {\rm NN};{\bf k}\rangle^{\rm in}$ is a 2-nucleon scattering state in the
centre-of-mass (COM) frame with nucleon momenta ${\bf k}$ and $-{\bf k}$ 
and total COM energy $W=2\sqrt{{\bf k}^2+m^2}$, $m$ is the nucleon mass
and $N({\bf x},t)$ is a local nucleon field operator. Both the nucleon field
operators and the 2-nucleon state depend on additional quantum numbers
(total spin $S$, isospin, etc.), which are suppressed in the above formula
for simplicity.

The reason to call the object defined by this formula a wave function is that
it can be shown that
at large nucleon separation ($r=\vert{\bf x}\vert\to\infty$) the interaction
between them can be neglected and it behaves like a free wave function:
\begin{equation}
(k^2+{\bf\nabla}^2)\psi^{\rm NBS}_{\bf k}({\bf x})\approx0,\qquad\quad
k=\vert{\bf k}\vert.
\end{equation}
Moreover, it can also be shown \cite{Aoki:2009ji,Ishizuka2009a} that
its radial component behaves for large separation $r$ as
\begin{equation}
\phi^{\rm NBS}_k(r;L,S)\approx \frac{\sin(kr-L\pi/2+\delta_{LS}(k))}{kr}
{\rm e}^{i\delta_{LS}(k)},
\end{equation}
where $L$ is the total angular momentum of the 2-nucleon state. Thus the 
exact scattering phase shifts $\delta_{LS}(k)$
are encoded in the NBS wave function. But
it contains much more information and motivated by the above
wave function interpretation one can define the NBS potential by writing
\begin{equation}
(E_{\bf k}-H_o)\psi^{\rm NBS}_{\bf k}({\bf x})=U^{\rm NBS}_{\bf k}({\bf x})
\psi^{\rm NBS}_{\bf k},
\end{equation}
where
\begin{equation}
E_{\bf k}=\frac{{\bf k}^2}{2M},\qquad
H_o=-\frac{1}{2M}{\bf\nabla}^2
\end{equation}
and $M$ is the reduced mass $M=m/2$. This resembles the non-relativistic 
Schr\"odinger equation with potential $U^{\rm NBS}_{\bf k}({\bf x})$.
Indeed, the lattice measurements found that $U^{\rm NBS}_{\bf k}({\bf x})$
is very similar to the phenomenological nuclear potential. At large distance 
it has an attractive tail, but at shorter distances it develops a
characteristic repulsive core (RC). While the long distance attraction has
long been understood by nuclear theorists and it is due to meson exchanges, it
was the first time that the RC has been obtained from a first principles 
calculation. 

Later the same method has been successfully applied also to other
hadronic interactions: this included the baryon-baryon potential
\cite{Inoue:2010hs,Inoue:2010es} and the study of 3-body nuclear forces
\cite{Doi:2011gq}. Short distance behaviour of the NBS wave function and 
potential can be analytically studied, thanks to the asymptotic freedom 
property of QCD, by operator product expansion and renormalization group
techniques \cite{Aoki:2010kx,Aoki:2009pi,Aoki:2010uz,Aoki:2011aa,Aoki:2012xa}.

Despite these successes, there are also some serious open problems within 
this approach. First, the wave function depends on the choice of the 
interpolating field $N({\bf x},t)$ used for nucleons. While in lattice studies
$N({\bf x},t)$ was naturally represented by a local, gauge invariant 3-quark 
operator, it is not known to what extent the resulting NBS potential
depends on this choice. Secondly, unlike the potential term in the 
Schr\"odinger equation, $U^{\rm NBS}_{\bf k}({\bf x})$ is energy (momentum)
dependent due to the relativistic nature of the problem. A possible solution
of this problem is to define \cite{Aoki:2008hh,Aoki:2009ji} a new, non-local, 
but energy independent potential operator. This non-local operator can be
approximated by a series containing terms with derivative operators
of increasing power. The leading term is a local potential and it is again 
similar to the phenomenological potential. Alternatively, since the energy 
dependence is weak at low energies, one can define the zero-momentum potential
\begin{equation}
U_o({\bf x})=\lim_{{\bf k}\to0} U^{\rm NBS}_{\bf k}({\bf x}).
\end{equation}
It can be shown \cite{Aoki:2008yw} that $U_o$ correctly reproduces the 
scattering lengths, but already the next to leading order parameter for low 
energy scattering, the effective range, may differ from the true one.

The problem of energy dependence has been studied in some $1+1$ dimensional
integrable field theory models \cite{Aoki:2008yw}, where the NBS wave function
can be represented by the form factor expansion. In these studies the Ising 
model and the O$(3)$ nonlinear sigma model were considered and it was found 
that $U_o({\bf x})$ is indeed a good approximation at low energies where 
the energy dependence is weak.

A more interesting toy model to study would be the sine-Gordon (SG) model, 
because unlike in the Ising model and the O$(3)$ model (which are free and 
repulsive, respectively), here we have both repulsive (soliton-soliton)
and attractive (soliton-antisoliton) scattering and in addition there are 
soliton-antisoliton bound states (breathers). The form factors are in
principle available also for this model, but to construct the NBS wave 
function via the form factor expansion would be very involved technically.
Luckily, an alternative description of the SG model exists since it is known 
that for any fixed particle number subspace of the SG field theory Hilbert
space  there is a corresponding Ruijsenaars-Schneider (RS) type relativistic
quantum mechanical description \cite{RS1}. The RS wave function is  
known \cite{Ruijsenaars:2000hg} for both soliton-soliton and 
soliton-antisoliton scattering and exactly reproduces the scattering phase 
shifts of SG field theory. Moreover, also the soliton-antisoliton bound state
spectrum is calculable and exactly match the SG results.

In this paper we take one more backward step and consider the classical
relativistic RS 2-particle scattering problem. Energy dependence of the 
potential is already present in this system but here the problem can be 
completely solved using textbook results for classical inverse scattering.
We can find the relation between the zero-momentum potential and 
the true effective potential analytically. One can hope that the zero-momentum potential
versus effective potential relation can similarly be found in the relativistic
RS quantum mechanical problem, using the existing methods of quantum inverse
scattering.

The paper is organized as follows. 
In section 2 we review the RS type relativistic 2-particle mechanics. 
In section 3 we construct the effective potential using classical inverse 
scattering, which is described in detail (adapted to and generalized for
our problem) in the appendix of the paper. We give our conclusions in
section 4.

\section{Ruijsenaars-Schneider type 2-particle problem}

Ruijsenaars-Schneider type models are a particular realization
of the Hamiltonian construction of relativistic point particle interaction in $1+1$ dimension.
The starting point for the latter is the relativistic phase space spanned by the canonical
variables $q_a$, $\theta_b$ satisfying 
\begin{equation}
\{q_a,q_b\}=\{\theta_a,\theta_b\}=0,\qquad\quad
\{q_a,\theta_b\}=\delta_{ab},\qquad a,b=1,2\dots,N.
\end{equation} 
For relativistic invariance we have to construct the three generators of the $1+1$ dimensional
Poincar\'e group, the Hamiltonian ${\cal H}$, the momentum ${\cal P}$, and
the Lorentz-boost ${\cal K}$, which satisfy the Poisson-bracket relations
\begin{equation}
\{{\cal H},{\cal P}\}=0,\qquad\quad
\{{\cal H},{\cal K}\}={\cal P},\qquad\quad
\{{\cal P},{\cal K}\}=\frac{1}{c^2}{\cal H}.
\label{Poin}
\end{equation} 
Using the Hamiltonian vector fields $\hat{\cal H}$ and $\hat{\cal P}$ associated with 
${\cal H}$ and ${\cal P}$ respectively, we can calculate
the time and space derivatives of any phase space function ${\cal F}$ by the usual formulas
\begin{equation} 
\hat{\cal H}{\cal F}=\{{\cal H},{\cal F}\}=\dot{\cal F},\qquad\quad
\hat{\cal P}{\cal F}=\{{\cal P},{\cal F}\}={\cal F}^\prime.
\end{equation} 
Further we can calculate the time and space \lq\lq flows" of the canonical coordinates by
solving the differential equations 
\begin{equation}
\frac{\partial}{\partial t}Q_a(t;q,\theta)=\dot q_a(Q,T),\qquad
\frac{\partial}{\partial t}T_b(t;q,\theta)=\dot \theta_b(Q,T)
\end{equation} 
with initial conditions
\begin{equation}
Q_a(0;q,\theta)=q_a,\qquad\quad T_b(0;q,\theta)=\theta_b
\end{equation} 
for the time flow $Q_a(t;q,\theta)$, $T_b(t;q,\theta)$ and
\begin{equation}
\frac{\partial}{\partial x}\bar Q_a(x;q,\theta)=q_a^\prime(Q,T),\qquad
\frac{\partial}{\partial x}\bar T_b(x;q,\theta)=\theta^\prime_b(Q,T)
\end{equation} 
with initial conditions
\begin{equation}
\bar Q_a(0;q,\theta)=q_a,\qquad\quad \bar T_b(0;q,\theta)=\theta_b
\end{equation} 
for the space flow $\bar Q_a(t;q,\theta)$, $\bar T_b(t;q,\theta)$.

The final step is finding the physical particle coordinates $x_a(q,\theta)$, $a=1,2,\dots,N$
as functions of the phase space variables. The construction we are using here is explained
in~\cite{Balog2014} and is based on $N$ Lorentz-invariant (not Poincar\'e invariant!)
phase space functions $\rho_a(q,\theta)$,
\begin{equation}
\hat{\cal K}\rho_a=\{{\cal K},\rho_a\}=0,\qquad\quad a=1,2,\dots,N.
\end{equation} 
Given $\rho_a$, we can calculate its space flow
\begin{equation}
R_a(x;q,\theta)=\rho_a(\bar Q,\bar T)
\end{equation} 
and the trajectory variable (coordinate) of the $a^{\rm th}$ particle is defined by the
implicit equation
\begin{equation}
R_a(x_a;q,\theta)=0.
\end{equation} 
Finally the time-dependent trajectory is given by
\begin{equation}
x_a(t;q,\theta)=x_a(Q,T).
\end{equation} 

The Ruijsenaars-Schneider Ansatz \cite{RS1} for two particles is of the form 
\begin{equation}
{\cal H}=mc^2(\cosh\theta_1+\cosh\theta_2)f(q_1-q_2),\qquad
{\cal P}=mc(\sinh\theta_1+\sinh\theta_2)f(q_1-q_2),
\end{equation} 

\begin{equation}
{\cal K}=-\frac{1}{c}(q_1+q_2),
\end{equation} 
where $m$ is the mass of the particles and
$f(q)$ is an even, positive real function, which we can parametrize as
\begin{equation}
f^2(q)=1+b(q).
\end{equation} 
$b(q)$ is, as we will see, the zero-momentum potential (up to rescaling). 
It is easy to check that the relations (\ref{Poin}) are satisfied for any\footnote{This is
no longer true for more than two particles, see \cite{RS1}.} such $f(q)$.

The best known examples are of hyperbolic type,
\begin{equation}
b(q)=\frac{\gamma^2}{\sinh^2(\omega q)}\qquad{\rm and}\qquad
b(q)=-\frac{\gamma^2}{\cosh^2(\omega q)}.
\end{equation} 
The inverse $\sinh^2$ potential is monotonically repulsive (MR, see \ref{a2})
whereas the negative inverse $\cosh^2$ potential is of LA type (see \ref{a4}).
The constant $\omega$ can be written as $1/mc\ell$ where $\ell$ is a length scale, and
the dimensionless coupling constant $\gamma$ is restricted in the LA case by $\gamma\leq1$.
The Sine-Gordon model corresponds to the choice $\gamma=1$ \cite{RS1}.

For the construction of the trajectory variables we can use \cite{RS1}
\begin{equation}
\rho_a(q,\theta)=q_a,\qquad\quad R_a(x;q,\theta)=\bar Q_a(x;q,\theta). 
\end{equation}
It turns out to be useful to introduce the centre-of-mass and relative coordinates
and momenta
\begin{equation}
\zeta=q_1+q_2,\qquad q=q_1-q_2;\qquad\quad 2\tau=\theta_1+\theta_2,\qquad
2u=\theta_1-\theta_2.
\end{equation} 
In terms of these,
\begin{equation}
{\cal H}= 2mc^2\varepsilon\cosh\tau,\qquad\quad
{\cal P}= 2mc\varepsilon\sinh\tau,
\end{equation} 
which shows that
\begin{equation}
\varepsilon=f(q)\cosh u
\end{equation} 
is the (Poincar\'e invariant) total mass, normalized to $1$, and the meaning of $\tau$
is the rapidity of the COM of the 2-particle system.

It is easy to see that
\begin{equation}
\dot\tau=0\qquad\quad{\rm and}\qquad\quad \dot\zeta=-2mc\varepsilon\sinh\tau,
\end{equation} 
thus it is consistent to go to the COM system $\tau=\zeta=0$. 
This simplifies the construction of the trajectory variables enormously and we find that
in the COM system
\begin{equation}
x_1=-x_2=\frac{q}{2mc\varepsilon}.
\end{equation} 
For the remaining
relative variables $q$, $u$ we introduce the corresponding time flows $Q$, $U$.
We also introduce the relative physical coordinate
\begin{equation}
y(t)=x_1(t)-x_2(t)=\frac{Q}{mc\varepsilon}.
\end{equation} 

The COM dynamics of the 2-particle Ruijenaars-Schneider model is equivalent to the
conservation law
\begin{equation}
\frac{1}{4}\dot y^2+\frac{1}{\varepsilon^2}W_o(\varepsilon y)=H^{\rm NR}={\rm const.},
\label{conserv}
\end{equation} 
where
\begin{equation}
W_o(x)=c^2b(mcx)
\end{equation}  
is the zero-momentum potential. The energy constant is given by
\begin{equation}
H^{\rm NR}=c^2\left(1-\frac{1}{\varepsilon^2}\right).
\end{equation} 
For scattering states of asymptotic velocity $v$ (in the COM system), where
\begin{equation}
\varepsilon=\frac{1}{\sqrt{1-\frac{v^2}{c^2}}}
\end{equation} 
we have
\begin{equation}
H^{\rm NR}=v^2,
\end{equation} 
whereas for bound states of mass $m_{\rm B}$ 
(where $\varepsilon=m_{\rm B}/2m$) we can use the parametrization
\begin{equation}
m_{\rm B}=2m-\frac{mh}{c^2} \qquad\quad (0\leq h\leq2c^2)
\end{equation}  
and we find
\begin{equation}
H^{\rm NR}=-h\frac{1-\frac{h}{4c^2}}{\left(1-\frac{h}{2c^2}\right)^2}.
\end{equation} 

(\ref{conserv}) looks like a non-relativistic problem, except for rescaling with
the state-dependent constant of motion $\varepsilon$. The corresponding NR problem is
\begin{equation}
\frac{1}{4}\dot z^2+W_o(z)=H_o={\rm const.},
\label{conservo}
\end{equation} 
for the NR variable $z(t)$. (\ref{conserv}) and (\ref{conservo}) coincide for $v=0$, which
justifies the name zero-momentum potential for $W_o$.

For the NR problem the energy constant can be written
\begin{equation}
H_o=v_o^2\quad{\rm (scattering)},\qquad\quad
H_o=-h_o\quad{\rm (bound\ state\ problem)},\quad 0\leq h_o\leq b_o\leq c^2.
\end{equation} 
(Here $-b_o$ is the minimum of $W_o$.)

The solution of the physical problem (\ref{conserv}) is obtained from the solution of the
fictious NR problenm (\ref{conservo}) by putting
\begin{equation}
y(t)=\frac{1}{\varepsilon}z(t)
\end{equation} 
and choosing
\begin{equation}
H_o=\varepsilon^2 H^{\rm NR}.
\end{equation} 
This corresponds to the choice
\begin{equation} 
v_o=\varepsilon v\quad{\rm (scattering)},\qquad\quad
h_o=h\left(1-\frac{h}{4c^2}\right)\quad{\rm (bound\ state\ problem)}.
\end{equation}

\section{Effective potential}

The following discussion is based on the theory of classical inverse scattering described
in appendix A. 

Taking into account the $\varepsilon$ dependence of the physical problem and the scaling
rules of \ref{a7} we see that the
physical (relativistic) scattering data are simply related to the ones calculated in
the NR problem:
\begin{equation}
X^{{\rm rel}}(v)=\frac{1}{\varepsilon}X_o(\varepsilon v),\qquad\quad
P^{{\rm rel}}(h)=P_o\left(h\left(1-\frac{h}{4c^2}\right)\right).
\end{equation} 
Here $P$ is the period in case of bound motion and $X(v)=-vT(v)$ is the displacement 
corresponding to the time delay $T(v)$. The time delay is the classical counterpart
of the quantum phase shift. It is the energy derivative of the phase shift in the
semiclassical ($\hbar\to0$) limit.
The formula for the displacement becomes especially simple if we introduce the
(mass-reduced) momentum variable $q$,
\begin{equation}
p=mq,\qquad\qquad q=\frac{v}{\sqrt{1-\frac{v^2}{c^2}}}.
\end{equation} 
We denote the displacement as funtion of this momentum variable by $\tilde X^{{\rm rel}}$
and we get
\begin{equation}
\tilde X^{{\rm rel}}(q)=\frac{1}{\sqrt{1+\frac{q^2}{c^2}}}X_o(q).
\end{equation} 
In the bound state problem
\begin{equation}
0\leq h\leq b_{{\rm rel}},
\end{equation} 
where
\begin{equation}
b_{{\rm rel}}-\frac{b^2_{{\rm rel}}}{4c^2}=b_o\leq c^2.
\end{equation} 

For the Sine-Gordon model soliton-soliton scattering we have to take as zero-momentum
potential our $1/\sinh^2$  MR example (\ref{MRex}) with $g=c$ and we find
\begin{equation}
\tilde X^{{\rm rel}}(q)=\frac{\ell}{2}\frac{1}{\sqrt{1+\frac{q^2}{c^2}}}\ln\left(1+\frac{c^2}
{q^2}\right).
\label{SGss}
\end{equation} 

The Sine-Gordon soliton-antisoliton scattering corresponds to the zero-momentum
potential $-1/\cosh^2$  in our LA example (\ref{LAex}) with $g=c$ and as shown in \ref{a9}
the scattering
displacement formula is exactly the same as (\ref{SGss}). For the relativistic period we find
\begin{equation}
P^{{\rm rel}}(h)=\frac{\ell\pi}{\sqrt{h}}\frac{1}{\sqrt{1-\frac{h}{4c^2}}},\qquad\quad
0\leq h\leq 2c^2.
\end{equation}

\begin{figure}
\begin{flushleft}
\leavevmode
\centerline{\includegraphics[width=10cm]{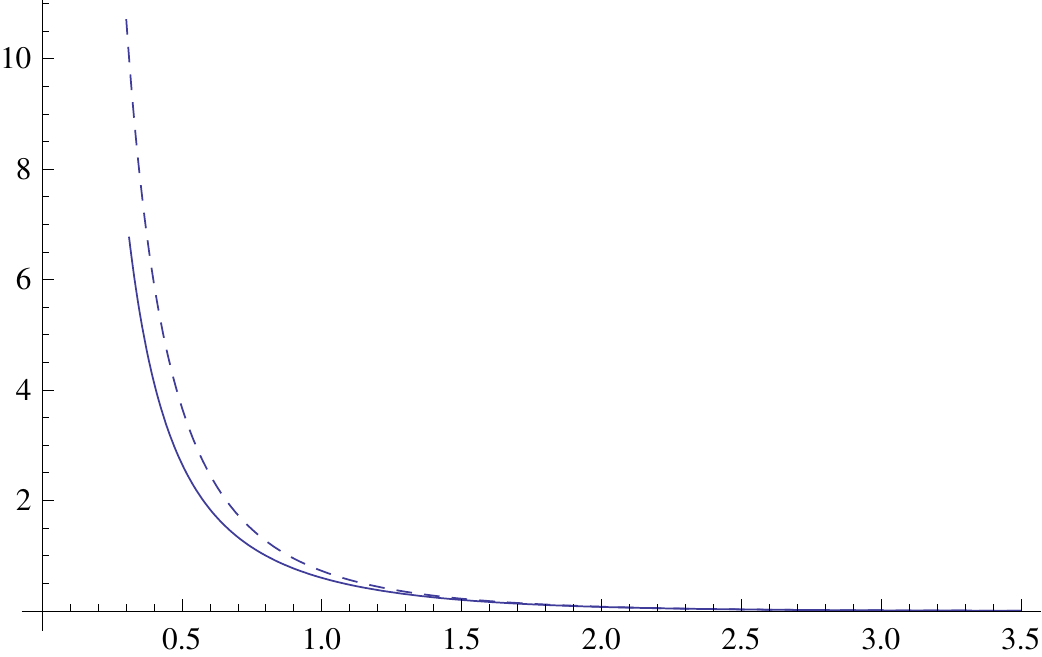} }
\end{flushleft}
\caption{{\footnotesize
Sine-Gordon effective potential (solid line). The dashed line is the corresponding 
zero-momentum potential. The plots show $W^{\rm eff}/mc^2$ vs. $x/\ell$.
}}
\label{figA}
\end{figure}

Since the relativistic and NR scattering data are very similar, the following question
arises naturally. Is there a NR effective potential $W^{{\rm eff}}$ such that the
physical, relativistic scattering data (in the COM frame) are exactly reproduced 
by using a non-relativistic Hamiltonian with potential $W^{{\rm eff}}$? In other words,
we require that
\begin{equation}
\tilde X^{{\rm rel}}(q)=X^{{\rm eff}}(q),\qquad\quad
P^{{\rm rel}}(h)=P^{{\rm eff}}(h),\quad 0\leq h\leq b_{{\rm rel}},\quad b_{{\rm rel}}=
b_{{\rm eff}}.
\end{equation} 

For the SG soliton-soliton scattering, the answer is yes. We simply take the physical
result (\ref{SGss}) and use the formulas given in \ref{a8} to obtain the effective potential
by using the techniques of classical inverse scattering for MR type potentials. The 
effective potential is given by an integral formula. The integral
cannot be calculated analytically, but it is easily obtained by 
numerical integration. The result is shown in Fig. \ref{figA}. From the low energy
asymptotics of (\ref{SGss}) we can read of the parameters (see \ref{a10})
\begin{equation}  
{\cal L}=\frac{\ell}{2},\qquad u_o=2c,\qquad \hat\alpha=\frac{3\ell}{2},\qquad
\hat\beta=\frac{\ell}{2}
\end{equation} 
and using the results of \ref{a10} we can determine the large distance asymptotics of
the effective potential:
\begin{equation}
U^{{\rm eff}}(x)\approx 4mc^2{\rm e}^{-2x/\ell}\left\{1+\left(3-\frac{2x}{\ell}\right)
{\rm e}^{-2x/\ell}+\dots\right\}
\end{equation} 
The leading term is the same as for the zero-momentum potential, but the subleading
terms differ.

For the Sine-Gordon soliton-antisoliton problem the answer is no. As shown in \ref{a4} 
for LA type NR potentials there is a constraint between the scattering and bound state
data and in this case the constraint (\ref{constraint1}) between $\tilde X^{{\rm rel}}(q)$ and
$P^{{\rm rel}}(h)$ is not satisfied. Therefore no $W^{{\rm eff}}(x)$ exists.

\begin{figure}
\begin{flushleft}
\leavevmode
\centerline{\includegraphics[width=10cm]{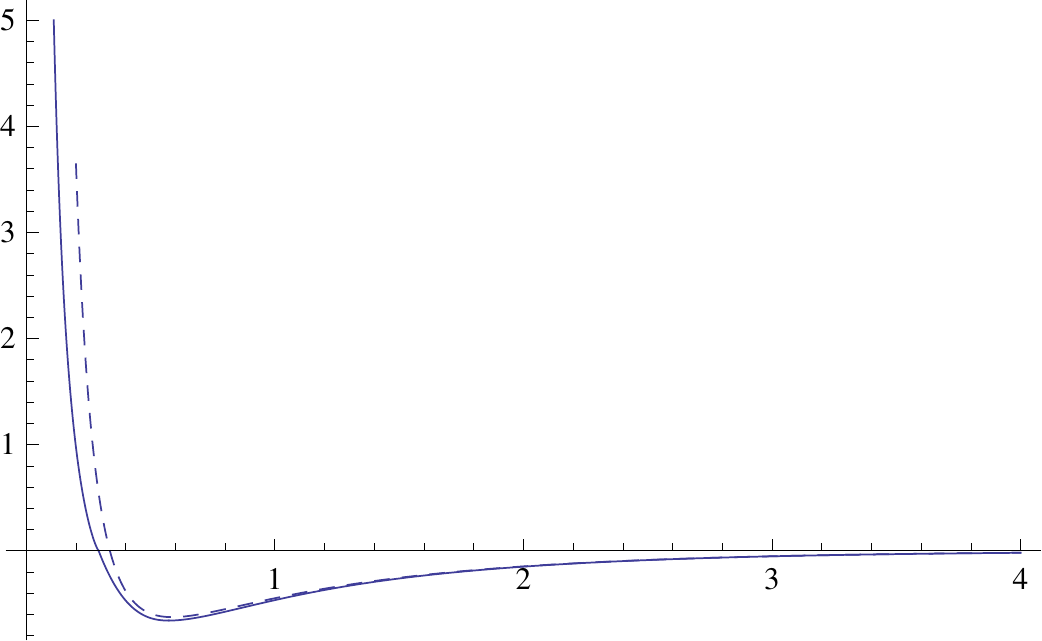} }
\end{flushleft}
\caption{{\footnotesize
Effective potential for an RC type potential with parameters $\beta=0.3$ 
and $\xi=1.4$ (solid line). The dashed line is the corresponding zero-momentum potential.
The plots show $W^{\rm eff}/mc^2$ vs. $x/\ell$.
}}
\label{figB}
\end{figure}

\begin{figure}
\begin{flushleft}
\leavevmode
\centerline{\includegraphics[width=10cm]{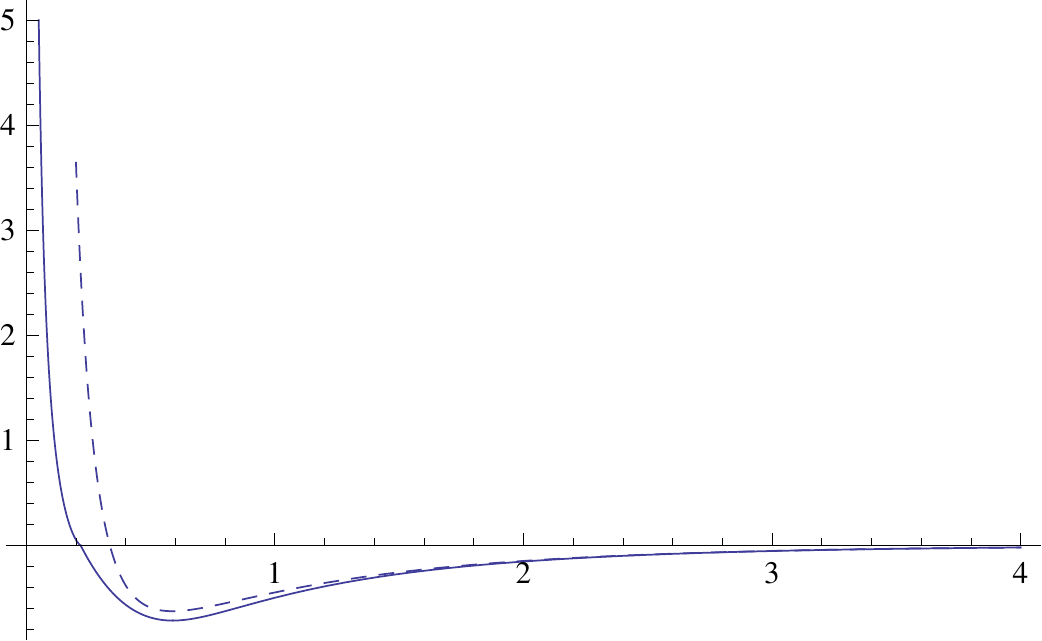} }
\end{flushleft}
\caption{{\footnotesize
Effective potential for an RC type potential with parameters $\beta=0.7$ 
and $\xi=1.4$ (solid line). The dashed line is the corresponding zero-momentum potential.
The plots show $W^{\rm eff}/mc^2$ vs. $x/\ell$.
}}
\label{figC}
\end{figure}

For our RC example (see \ref{a3}, \ref{a9}) the answer is again yes. We have to use both 
$\tilde X^{{\rm rel}}(q)$ and $P^{{\rm rel}}(h)$ to determine the two partial
inverse functions, which are then used to reconstruct $W^{{\rm eff}}(x)$.
We did this numerically. The results are shown in Figs. \ref{figB},\ref{figC},
for $\xi=1.4$ and the parameter values $\beta=B/c^2=0.3$, 
$\beta=B/c^2=0.7$ respectively.

\section{Conclusion}

The Nambu-Bethe-Salpeter potential as measured by the HAL QCD collaboration
can be identified, at low energies, with the zero-momentum nucleon potential.
This can be compared to the phenomenological nuclear potential, which has been
constructed to reproduce the nucleon scattering data (at low energies, below
the pion production threshold). This problem can be modelled in a $1+1$ 
dimensional toy model, the Sine-Gordon field theory. For the 2-particle case,
one can study the equivalent quantum mechanical problem, the relativistic
Ruijsenaars-Schneider model for two particles. In this paper we worked out the
zero-momentum potential $\to$ effective potential mapping in the semiclassical
limit of the RS model, using classical inverse scattering techniques. 
It turned out that the very existence of such a mapping depends crucially  
on the qualitative features of the potential. For repulsive scattering and 
potentials with a repulsive core, the zero-momentum and effective potentials 
are qualitatively very similar and quantitatively close at low energies.
The first one can be used to describe soliton-soliton scattering in the SG 
model and the second one is a $1+1$ dimensional model of the nucleon potential.
On the other hand, no such mapping exists for soliton-antisoliton scattering
and bound states in the SG model.

It is likely that quantum inverse scattering can be applied to study the same 
questions at the quantum mechanical level in SG/RS theory. Whether the
zero-momentum potential $\to$ effective potential mapping exists in the 
physically relevant $3+1$ dimensional nucleon problem is an open question.

\vspace{5ex}
\begin{center}
{\large\bf Acknowledgments}
\end{center}

This investigation was supported by the Hungarian National Science Fund OTKA
(under K83267). J.~B. would like to thank the CAS
Institute of Modern Physics, Lanzhou, where most of this work has been 
carried out, for hospitality.

\par\bigskip

\appendix

\section{Classical inverse scattering}

In this appendix we summarize the techniques used for classical inverse 
scattering.

\subsection{Landau-Lifshitz formula}
\label{a1}

\begin{figure}
\begin{flushleft}
\leavevmode
\centerline{\includegraphics[width=10cm]{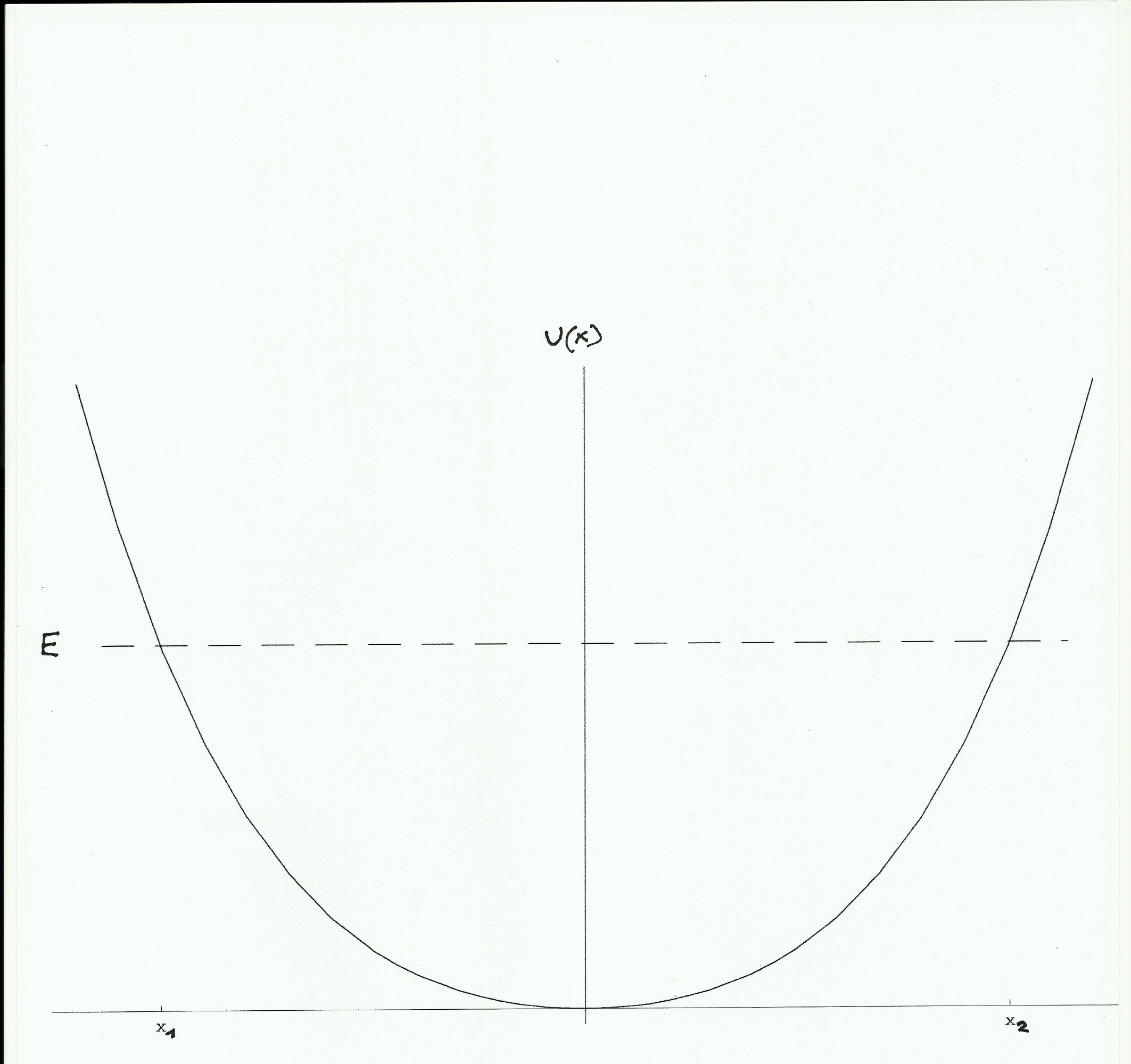} }
\end{flushleft}
\caption{{\footnotesize
Symmetric potential. The half-period of periodic motion with energy $E$
is the time the particle needs to move between the turning points $x_1$
and $x_2$.
}}
\label{fig1}
\end{figure}

A basic problem in analytic classical mechanics is to reconstruct the potential
for a point particle in one dimension if the period of oscillations for 
the bound motions as function of the energy is known. 
The solution of this problem can
be found in the book of Landau \& Lifshitz \cite{LL}. We take, for simplicity,
a symmetric potential $U(x)$ with $U(0)=0$ which is monotonically increasing
for $0\leq x<\infty$ (see Fig. \ref{fig1}). The Landau-Lifshitz trick is to consider
instead of the potential its inverse function $\xi(U)$. For a given energy $E$,
the bound motion of the particle is confined to $x_1\leq x\leq x_2$, where
$x_2=-x_1=\xi(E)$.

The half-period of oscillations is easily expressed as
\begin{equation}
T(E)=\sqrt{2m}\int_0^{\xi(E)}\frac{{\rm d}x}{\sqrt{E-U(x)}}=
\sqrt{2m}\int_0^E\xi^\prime(U)\frac{{\rm d}U}{\sqrt{E-U}}.
\label{LL1}
\end{equation}
The trick is to change the integration variable to $U$. $m$ is the mass 
of the particle. Given $T(E)$, (\ref{LL1}) is an Abel-type linear 
integral equation for the unknown function $\xi(U)$. The solution is given
by the simple formula \cite{LL}
\begin{equation}  
\xi(U)=\frac{1}{\pi \sqrt{2m}}\int_0^U\frac{T(E){\rm d}E}{\sqrt{U-E}}.
\end{equation}

\subsection{Time delay in classical one-dimensional scattering, monotonic repulsive (MR) potential}
\label{a2}

\begin{figure}
\begin{flushleft}
\leavevmode
\centerline{\includegraphics[width=10cm]{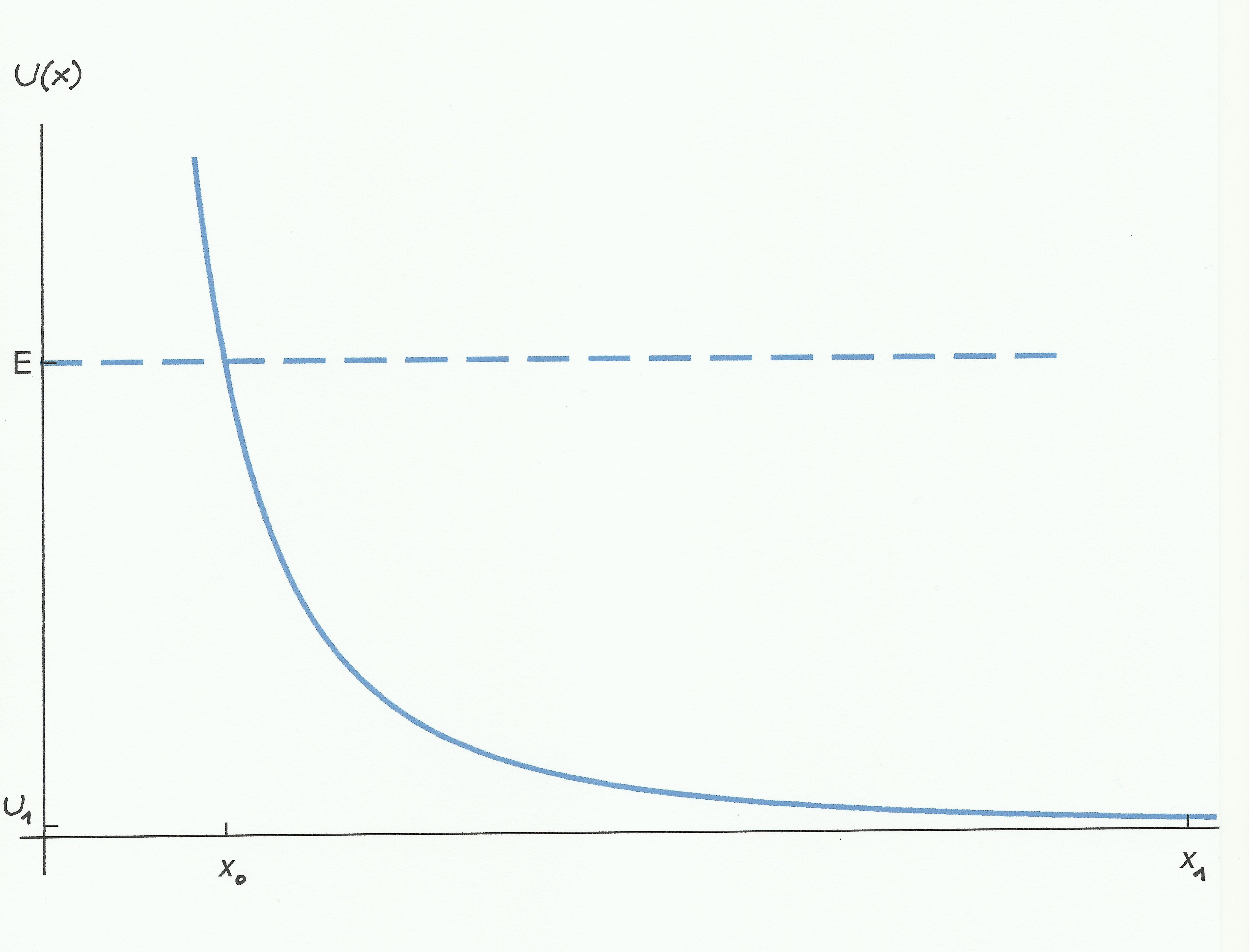} }
\end{flushleft}
\caption{{\footnotesize
Monotonically repulsive potential. The time delay is the time difference between
the actual motion from the turning point $x_o$ to $x_1$ and
the free motion from the origin to a distant point $x_1$, 
in the limit $x_1\to\infty$.
}}
\label{fig2}
\end{figure}

A similar, but somewhat more complicated problem is to reconstuct the 
one dimensional potential from the classical time delay in scattering problems.
The details of the computation strongly depend on the type of the potential.
We start with the simplest case of a monotonically decreasing, repulsive (MR)
potential (see Fig. \ref{fig2}). Assuming $U(x)>0$, $U(\infty)=0$ and $U^\prime(x)<0$,
we can find again the inverse function $\xi(U)$. We will consider a scattering
process with fixed energy $E$. The energy can be parametrized as
\begin{equation}
E=\frac{1}{2}mv^2,
\end{equation}
where $v$ is the asymptotic velocity of the particle. The scattering process
is infinite so first we calculate the time necessary to reach the point $x_1$
starting from the turning point $x_o=\xi(E)$:
\begin{equation}
\sqrt{\frac{m}{2}}\int_{x_o}^{x_1}\frac{{\rm d}x}{\sqrt{E-U(x)}}=
-\sqrt{\frac{m}{2}}\int_{U_1}^E\xi^\prime(U)\frac{{\rm d}U}{\sqrt{E-U}}.
\end{equation}
If the potential were not there, the particle would move freely with constant
velocity $v$ (except from bouncing back from the origin) and the time 
from $0$ to $x_1$  would be
\begin{equation}
\frac{x_1}{v}=\sqrt{\frac{m}{2E}}x_1.
\end{equation}
The time delay $\Delta(E)$ is the time difference between the actual motion 
and the free one in the limit $x_1\to\infty$ ($U_1\to0$):
\begin{equation}
\Delta(E)=-\sqrt{2m}\left\{\frac{\xi(E)}{\sqrt{E}}+\int_0^E\xi^\prime(U){\rm d}U
\left[\frac{1}{\sqrt{E-U}}-\frac{1}{\sqrt{E}}\right]\right\}.
\label{DeltaMR}
\end{equation}
The derivation of the above formula is valid if
\begin{equation}
\lim_{x\to\infty}x^2U(x)=0,
\end{equation}
i.e. if the potential vanishes sufficiently fast at infinity.

Although the formula (\ref{DeltaMR}) is more complicated than the one in the
previous subsection, the corresponding integral equation can be solved 
by the same trick with the result
\begin{equation}  
\xi(U)=-\frac{1}{\pi \sqrt{2m}}\int_0^U\frac{\Delta(E){\rm d}E}{\sqrt{U-E}}.
\end{equation}

\subsection{Potential with repulsive core (RC)}
\label{a3}

\begin{figure}
\begin{flushleft}
\leavevmode
\centerline{\includegraphics[width=10cm]{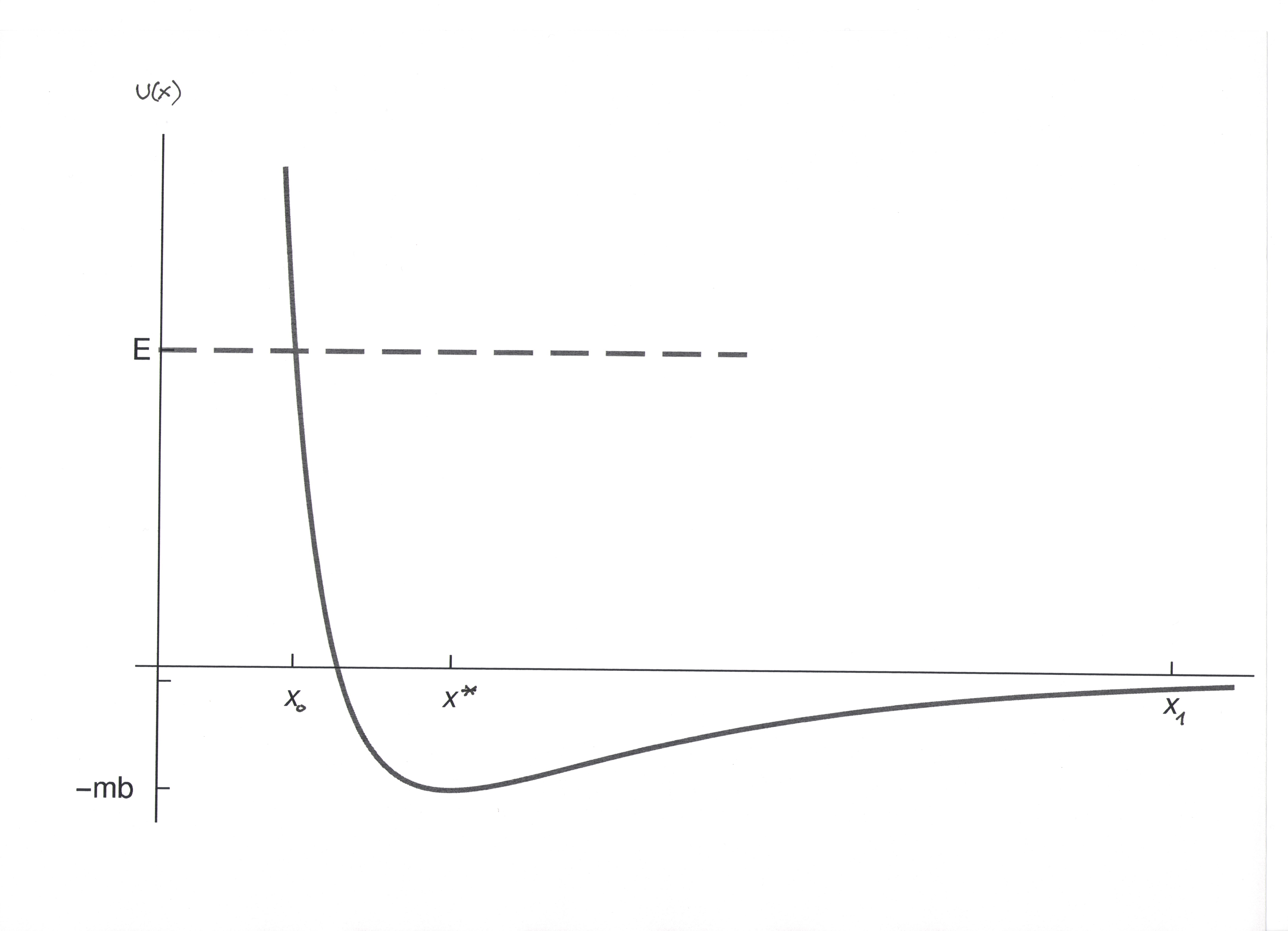} }
\end{flushleft}
\caption{{\footnotesize
Potential with repulsive core. The minimum of the potential ($-mb$) is at
$x=x^*$. For the calculation of the time delay the motion between the turning
point $x_o$ and a distant point $x_1$ is used.
}}
\label{fig3}
\end{figure}

The potential shown in Fig. \ref{fig3} is a one dimensional model of the nuclear 
potential. It consists of a monotonically decreasing part ($0<x<x^*$) and
a monotonically increasing part ($x^*<x<\infty$) with $U(\infty)=0$.
The minimum of the potential is at $x^*$ and it is parametrized as:
\begin{equation}
U(x^*)=-mb.
\end{equation}
Since there is no global inverse function, we have to use the two partial
functional inverse functions $\xi_1(U)$, $\xi_2(U)$. They are defined
for $\infty>U\geq -mb$ and $-mb\leq U<0$, respectively and satisfy
\begin{equation}
\xi_1(-mb)=\xi_2(-mb)=x^*.
\end{equation}
For motions with negative total energy $-mb<E<0$ it is useful to introduce
the \lq\lq width function"
\begin{equation}
d(V)=\xi_2(-V)-\xi_1(-V),\qquad\quad 0<V\leq mb.
\end{equation}

\begin{figure}
\begin{flushleft}
\leavevmode
\centerline{\includegraphics[width=10cm]{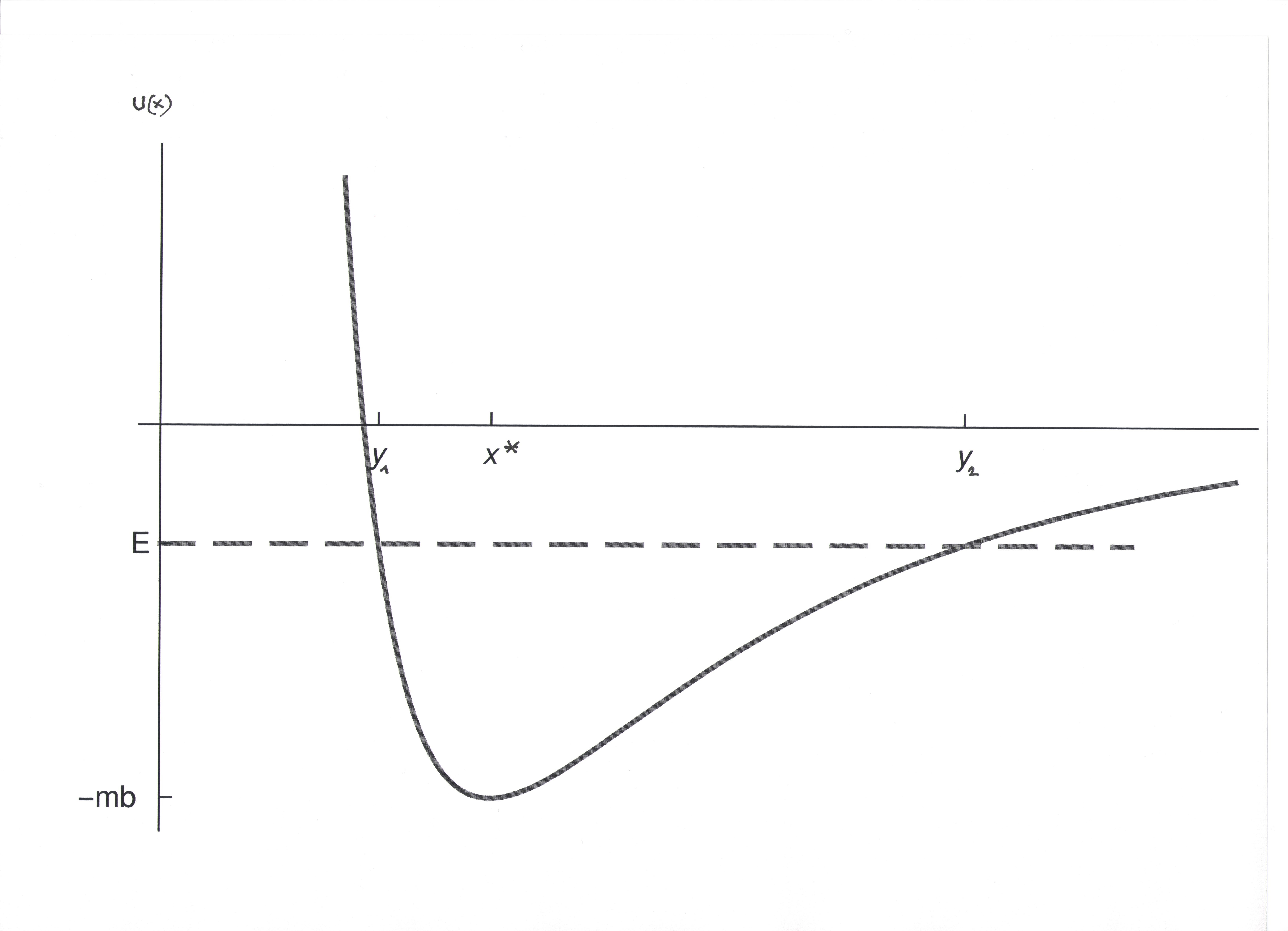} }
\end{flushleft}
\caption{{\footnotesize
Bound motion in an RC type potential. For negative total energy $E$
the turning points are $y_1$ and $y_2$.
}}
\label{fig4}
\end{figure}

The formula for the time delay for scattering processes is given by
\begin{equation}
\Delta(E)=-\sqrt{2m}\left\{\frac{\xi_1(E)}{\sqrt{E}}
+\int_0^E\xi_1^\prime(U){\rm d}U
\left[\frac{1}{\sqrt{E-U}}-\frac{1}{\sqrt{E}}\right]
+\int_0^{mb}d^\prime(V){\rm d}V
\left[\frac{1}{\sqrt{E+V}}-\frac{1}{\sqrt{E}}\right]
\right\}.
\label{DeltaRC}
\end{equation}
It depends on $\xi_1(U)$ and $d(V)$, i.e. on both inverse functions 
$\xi_1$, $\xi_2$. 
Using the LL trick, we can express $\xi_1(U)$ for $U\geq0$ as
\begin{equation}  
\xi_1(U)=-\frac{1}{\pi \sqrt{2m}}\int_0^U\frac{\Delta(E){\rm d}E}{\sqrt{U-E}}
-\frac{1}{\pi}\sqrt{U}\int_0^{mb}d(V)\frac{{\rm d}V}{\sqrt{V}(U+V)}.
\label{xi1RCpart}
\end{equation}
We see that this still depends on the width function. The scattering data alone
are not enough to find both inverse functions and reconstruct the potential.
For this we also need to consider the bound state problem (see Fig. \ref{fig4}).
First we have to calculate the half-period of periodic motions with
negative energy $E=-\varepsilon<0$:  
\begin{equation}
\tilde T(\varepsilon)=\sqrt{\frac{m}{2}}\int_{y_1}^{y_2}
\frac{{\rm d}x}{\sqrt{-\varepsilon-U(x)}}=
-\sqrt{\frac{m}{2}}\int_\varepsilon^{mb}d^\prime(V)\frac{{\rm d}V}
{\sqrt{V-\varepsilon}}.
\label{halfRC}
\end{equation}

Now we can use the LL-trick to determine the width function:
\begin{equation}  
d(V)=\frac{1}{\pi}\sqrt{\frac{2}{m}}\int_V^{mb}\frac{\tilde T(\varepsilon)
{\rm d}\varepsilon}{\sqrt{\varepsilon-V}}.
\end{equation}
Finally, using this result in (\ref{xi1RCpart}) we can reconstruct $\xi_1$ in terms
of scattering and bound state data:
\begin{equation}  
\xi_1(U)=-\frac{1}{\pi \sqrt{2m}}\int_0^U\frac{\Delta(E){\rm d}E}{\sqrt{U-E}}
-\frac{1}{\pi}\sqrt{\frac{2}{m}}\int_0^{mb}
\frac{\tilde T(\varepsilon){\rm d}\varepsilon}{\sqrt{\varepsilon+U}}.
\label{xi1RC}
\end{equation}

\subsection{Localized attractive potential (LA)}
\label{a4}

\begin{figure}
\begin{flushleft}
\leavevmode
\centerline{\includegraphics[width=10cm]{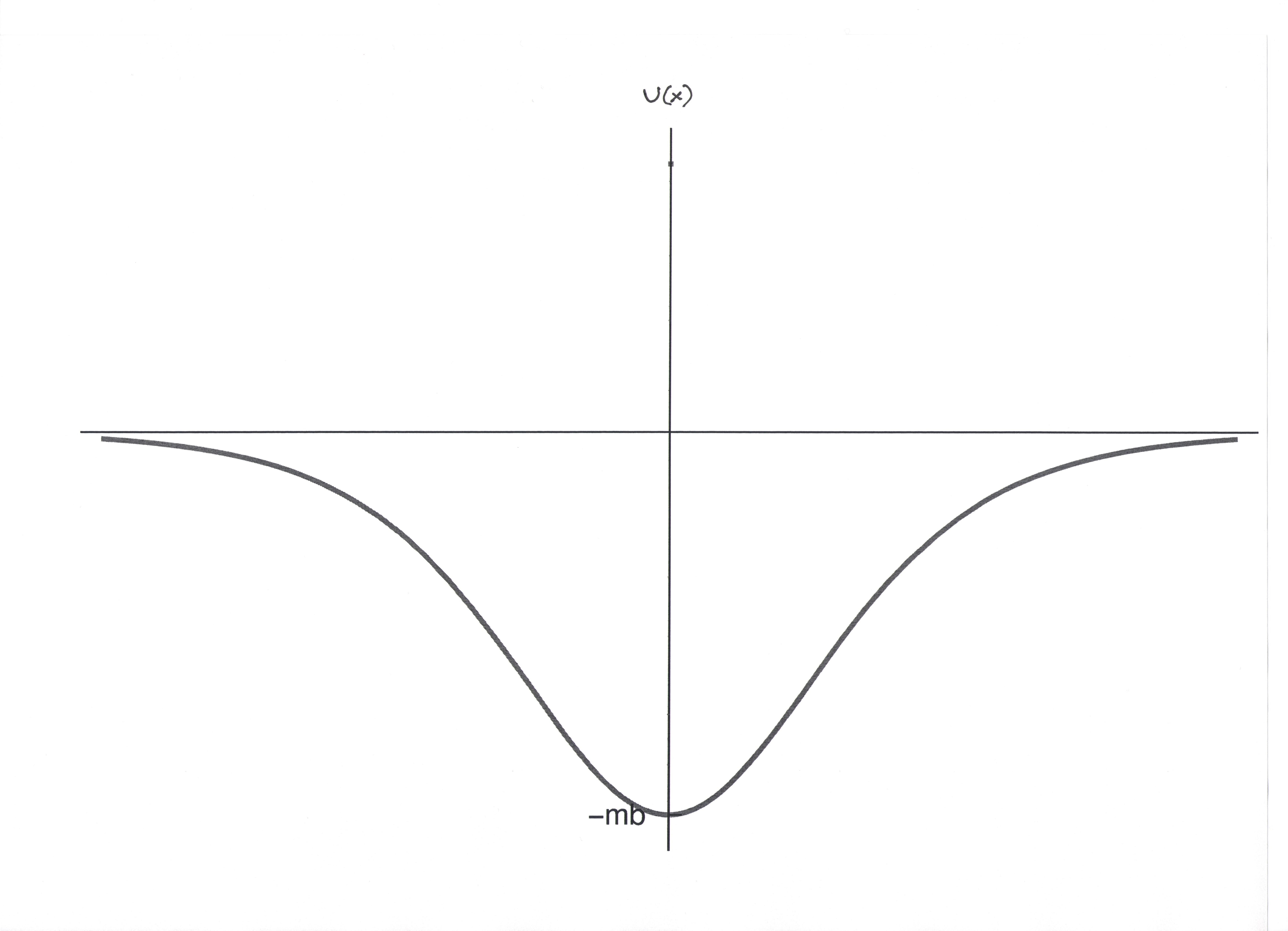} }
\end{flushleft}
\caption{{\footnotesize
Localized attractive potential. Both scattering and bound motions are possible.
}}
\label{fig5}
\end{figure}

The last example we discuss is shown in Fig \ref{fig5}. For simplicity, here we discuss a
symmetric, attractive potential, which takes its minimum value, $-mb$, 
at the origin. Here we can define the functional inverse $\xi(U)$
for $x\geq0$. We assume $U(\infty)=0$ again. There are scattering and bound motions
and we can calculate the time delay for $E>0$:
\begin{equation}
\Delta(E)=\sqrt{2m}\int_0^{mb}\xi^\prime(-V){\rm d}V
\left[\frac{1}{\sqrt{E+V}}-\frac{1}{\sqrt{E}}\right],
\label{DeltaLA}
\end{equation}
and also the half-period of bound motions for $E=-\varepsilon<0$:
\begin{equation}
\tilde T(\varepsilon)=
\sqrt{2m}\int_\varepsilon^{mb}\xi^\prime(-V)\frac{{\rm d}V}
{\sqrt{V-\varepsilon}}.
\label{halfLA}
\end{equation}
This last result is already enough to reconstruct the inverse potential by the
LL-trick:
\begin{equation}  
\xi(-V)=\frac{1}{\pi\sqrt{2m}}\int_V^{mb}\frac{\tilde T(\varepsilon)
{\rm d}\varepsilon}{\sqrt{\varepsilon-V}}.
\label{xiLA}
\end{equation}
The scattering time delay is determined by the same function and is not independent.
We find that there is a constraint between the time delay and the half-period:
\begin{equation}  
\Delta(E)=-\frac{1}{\pi\sqrt{E}}\int_0^{mb}\frac{\tilde T(\varepsilon)
\sqrt{\varepsilon}{\rm d}\varepsilon}{\varepsilon+E}.
\label{constraint}
\end{equation}

\subsection{Space-time picture of scattering}
\label{a5}

\begin{figure}
\begin{flushleft}
\leavevmode
\centerline{\includegraphics[width=10cm]{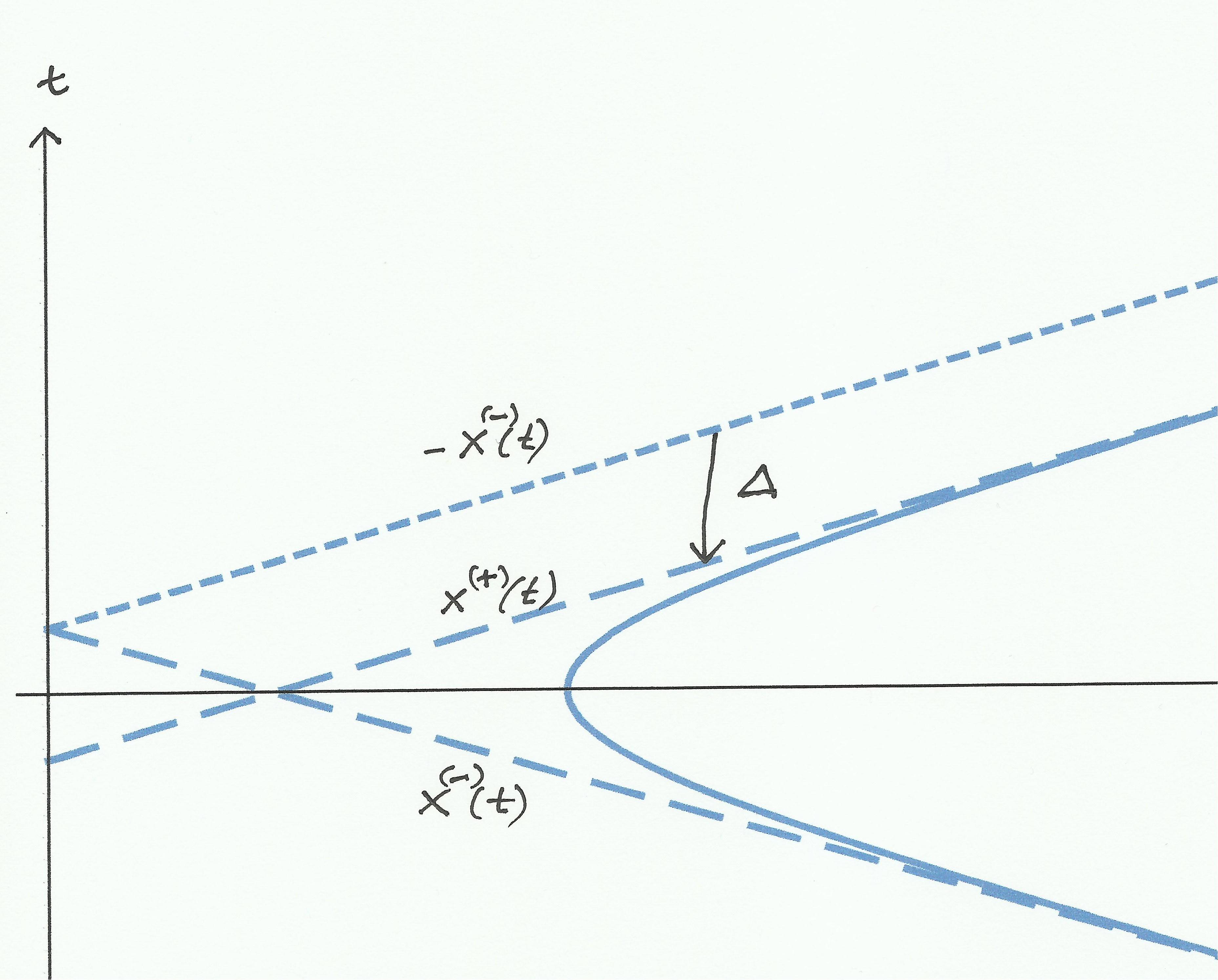} }
\end{flushleft}
\caption{{\footnotesize
Space-time diagram of a particle scattering off a potential. The time shift 
between the actual asymptotic motion and the free motion after bouncing back
at the origin is the time delay.
}}
\label{fig6}
\end{figure}

For repulsive scattering (MR and RC cases) the space-time diagram of the process
is depicted in Fig. \ref{fig6}. The free motion in the asymptotic past is given by
\begin{equation}
x(t)\approx x^{(-)}(t)=-vt+a,\qquad\qquad t\to -\infty
\end{equation}
and in the asymptotic future
\begin{equation}
x(t)\approx x^{(+)}(t)=vt+b,\qquad\qquad t\to +\infty.
\end{equation}
The values of the constants $a$, $b$ depend on the arbitrary choice of the origin
of the time coordinate, but their sum is uniquely determined by the asymptotic
velocity $v$, i.e. the energy of the process. An alternative definition of the time
delay is
\begin{equation}
x^{(+)}(t+\Delta)=-x^{(-)}(t).
\end{equation}
It is given by
\begin{equation}
\Delta=-\frac{a+b}{v}.
\end{equation}

Similarly, for the scattering process in the LA case 
\begin{equation}
x^{(-)}(t)=vt+a,\qquad\qquad 
x^{(+)}(t)=vt+b,\qquad\qquad 
\end{equation}
\begin{equation}
x^{(+)}(t+\Delta)=x^{(-)}(t),\qquad\qquad \Delta=\frac{a-b}{v}.
\end{equation}

\subsection{Two-particle problem}
\label{a6}

Let us scale out the mass $m$ from the one-particle problem introducing $W(x)$ by
\begin{equation}
U(x)=mW(x).
\end{equation}
Let us further introduce the notations
\begin{equation}
\tau([W];v)=\Delta,\qquad\qquad
T_o(h)=\tilde T(mh)\quad(E=-mh<0).
\end{equation}

\begin{figure}
\begin{flushleft} 
\leavevmode
\centerline{\includegraphics[width=10cm]{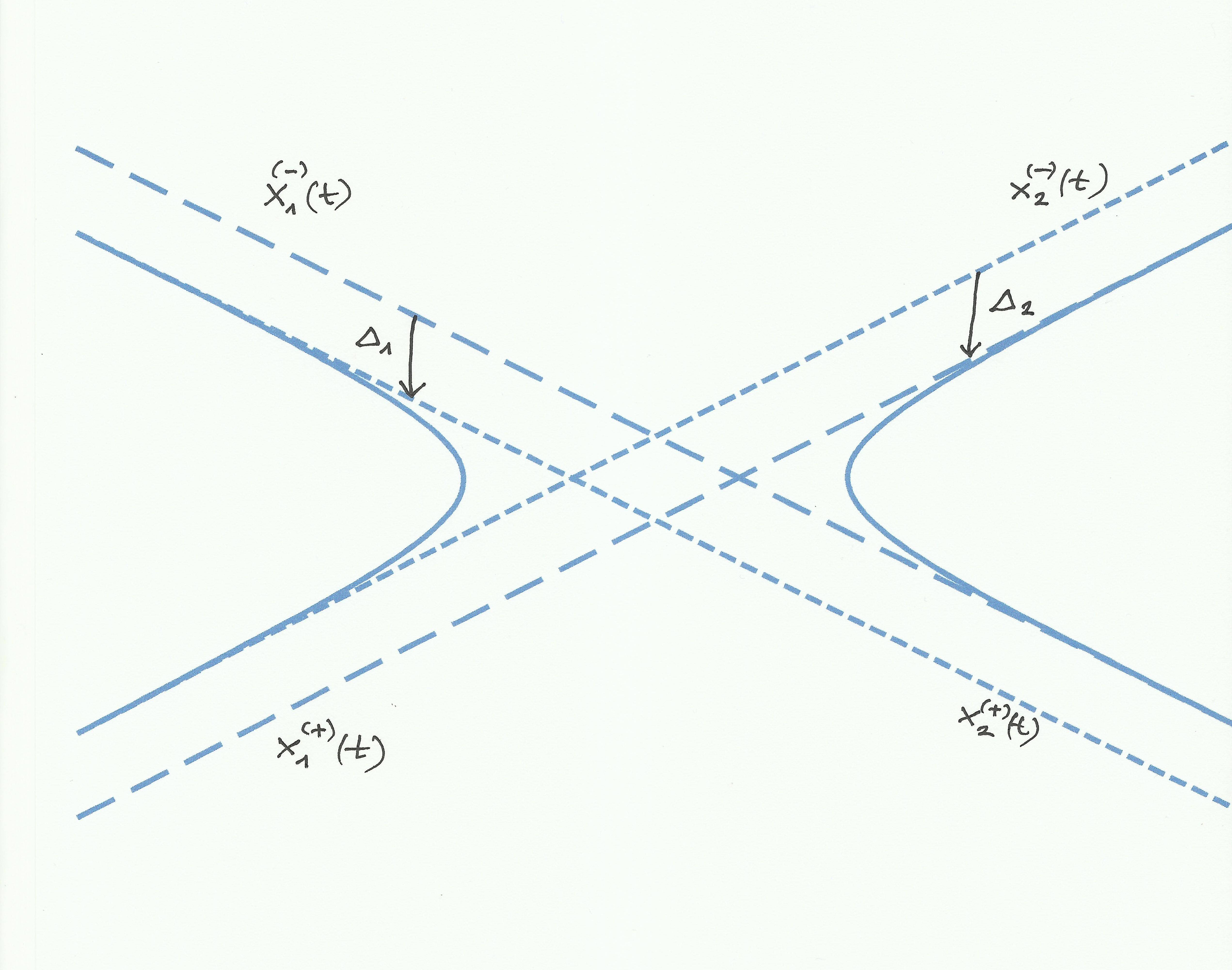} }
\end{flushleft}
\caption{{\footnotesize
Space-time diagram of a repulsive two-particle scattering process. The
time shift between the actual final asymptotics and the extrapolation
of the initial asymptotics is the time delay.
}}
\label{fig7}
\end{figure}

The simple exercises we have discussed in the previous subsections can be applied
to the study of 2-particle problems. Assuming that the particles are both of mass $m$
and interact through the potential $U(x_1-x_2)$, we can write down the equations
of motion:
\begin{equation}
m\ddot x_1=-U^\prime(x_1-x_2),\qquad\qquad
m\ddot x_2=U^\prime(x_1-x_2).
\end{equation}
As is well known, introducing the relative coordinate
\begin{equation}
y(t)=x_1(t)-x_2(t)
\end{equation}
we can reduce the problem to an effective 1-particle one
\begin{equation}
m\ddot y=-2U^\prime(y)
\end{equation}
with the same potential, but reduced mass $m/2$.

Let us now consider the case of repulsive scattering (Fig. \ref{fig7}). Because the 
scattering is elastic, the asymptotic velocities are swapped:
\begin{equation}
\begin{split}
t\to-\infty:\qquad\quad x_i(t)&\approx x_i^{(-)}(t)=v_it+a_i,
\qquad\qquad v_{\underline 1}=v_2,\\
t\to+\infty:\qquad\quad x_i(t)&\approx x_i^{(+)}(t)=v_{\underline i}t+b_i,
\qquad\qquad v_{\underline 2}=v_1.
\end{split}
\end{equation}
The time delays are determined by
\begin{equation}
x_2^{(+)}(t+\Delta_1)=x_1^{(-)}(t),\qquad\qquad \Delta_1=\frac{a_1-b_2}{v_1}
\end{equation}
and
\begin{equation}
x_1^{(+)}(t+\Delta_2)=x_2^{(-)}(t),\qquad\qquad \Delta_2=\frac{a_2-b_1}{v_2}.
\end{equation}
The kinematics is somewhat simplified in the COM frame. Here
$x_1(t)+x_2(t)=0$ and
\begin{equation}
-v_1=v_2=v,\qquad\qquad a_2=-a_1,\qquad b_2=-b_1.
\end{equation}
The time delays are equal:
\begin{equation}
\Delta_1=\Delta_2=-\frac{a_1+b_1}{v}=T(v).
\end{equation}
For the relative motion we have
\begin{equation}
y^{(-)}(t)=-2vt+2a_1,\qquad\quad
y^{(+)}(t)=2vt+2b_1,
\end{equation}
i.e. we have to consider an effective one-particle problem with (mass-reduced)
potential $2W$ and asymptotic velocity $2v$. We can calculate the time delay $\Delta$
in this effective problem and find
\begin{equation}
\Delta=\tau([2W];2v)=-\frac{2a_1+2b_1}{2v}=-\frac{a_1+b_1}{v}=T(v).
\end{equation}

The case of attractive scattering is very similar. We define
\begin{equation}
t\to\pm\infty\qquad\quad x_i(t)\approx x_i^{(\pm)}(t)=v_it+\left\{
\begin{aligned}&b_i\\&a_i\end{aligned}\right.
\end{equation}
and
\begin{equation}
x_i^{(+)}(t+\Delta_i)=x_i^{(-)}(t),\qquad\qquad \Delta_i=\frac{a_i-b_i}{v_i}.
\end{equation}
Again, in the COM frame the kinematics simplifies:
\begin{equation}
v_1=-v_2=v,\qquad a_2=-a_1,\quad b_2=-b_1,\qquad
\Delta_1=\Delta_2=\frac{a_1-b_1}{v}=T(v).
\end{equation}
For the effective one-particle problem we have
\begin{equation}
y^{(-)}(t)=2vt+2a_1,\qquad y^{(+)}(t)=2vt+2b_1
\end{equation}
and for the time delay $\Delta$
\begin{equation}
\Delta=\tau([2W];2v)=\frac{2a_1-2b_1}{2v}=\frac{a_1-b_1}{v}=T(v).
\end{equation}

\subsection{Scaling properties}
\label{a7}

Let us denote the solution of the equations of motion with (mass reduced) potential $W(x)$
by $x(t)$. It is the solution of
\begin{equation}
\ddot x(t)=-W^\prime(x(t)).
\end{equation}
If we rescale the time variable by a constant $\lambda$ we can define
\begin{equation}
z(t)=x(\lambda t).
\end{equation}
It solves
\begin{equation}
\ddot z(t)=-\lambda^2W^\prime(z(t)),
\end{equation}
i.e. it is the solution of the equations of motion with potential $\lambda^2 W(x)$.
We have seen that for repulsive scattering the asymptotics is given by
\begin{equation}
x^{(-)}(t)=-vt + a,\qquad\quad
x^{(+)}(t)=vt + b
\end{equation}
and the time delay is
\begin{equation}
\tau([W];v)=-\frac{a+b}{v}.
\end{equation}
After rescaling we have
\begin{equation}
z^{(-)}(t)=-v\lambda t + a,\qquad\quad
z^{(+)}(t)=v\lambda t + b
\end{equation}
and 
\begin{equation}
\tau([\lambda^2 W];\lambda v)=-\frac{a+b}{\lambda v}=\frac{1}{\lambda}\tau([W];v).
\end{equation}
The same scaling rule holds also for attractive scattering.

The time delay in the two-particle problem in the COM frame is
\begin{equation}
T(v)=\tau([2W];2v)=\frac{1}{\sqrt{2}}\tau([W];\sqrt{2}v).
\end{equation}
Here we have used the scaling rule with $\lambda=\sqrt{2}$, $v\rightarrow \sqrt{2}v$.

Later we will see that the formulas become simpler if we use instead of the 
time delay $T(v)$ the space displacement
\begin{equation}
X(v)=-vT(v).
\end{equation}
We have defined it with a minus sign because it turns out that in all our examples
the time delay is actually negative (which means that the interacting particles move
faster than the free ones).

For bound states in the original problem with half-period $T_o(h)$ we have
\begin{equation}
x(t+2T_o)=x(t).
\end{equation}
Here $-h$ is the conserved (mass-reduced) one-particle energy
\begin{equation}
-h=\frac{1}{2}\dot x^2(t)+W(x(t)).
\end{equation}
If we denote by $P$ the full period of the time-rescaled motion we have
\begin{equation}
z(t+P)=z(t).
\end{equation}
This gives $\lambda P=2T_o$ and for the two-particle case in the 
COM frame ($\lambda=\sqrt{2}$)
\begin{equation}
P=\sqrt{2}T_o.
\end{equation}
The (mass-reduced) two-particle energy is 
\begin{equation}
\frac{1}{2}\dot x_1^2(t)+\frac{1}{2}\dot x_2^2(t)+W(x_1(t)-x_2(t))=
\frac{1}{4}\dot y^2(t)+W(y(t))=\frac{1}{2}\dot x^2(\sqrt{2}t)+W(x(\sqrt{2}t))=-h,
\end{equation}
i.e. it is the same as the corresponding one-particle energy. Thus we have simply
\begin{equation}
P(h)=\sqrt{2}T_o(h)=\sqrt{2}\tilde T(mh),\qquad\quad E=-mh<0.
\end{equation}

\subsection{Simplified inverse formulas}
\label{a8}

Using the new variables $X(v)$ (displacement in the COM frame) and $P(h)$ (full period
of bound motion with total COM energy $E=-mh$) the inverse formulas are simplified
and can be written as follows.

MR type potential:
\begin{equation}
\xi(mW)=\frac{2}{\pi}\int_0^{\pi/2}X(\sqrt{W}\sin\varphi){\rm d}\varphi.
\end{equation}

RC type potential:
\begin{equation}
\xi_1(mW)=\frac{2}{\pi}\int_0^{\pi/2}X(\sqrt{W}\sin\varphi){\rm d}\varphi
-\frac{1}{\pi}\int_0^b\frac{P(h){\rm d}h}{\sqrt{h+W}},
\end{equation}
\begin{equation}
d(m{\cal V})=\frac{1}{\pi}\int_{\cal V}^b\frac{P(h){\rm d}h}{\sqrt{h-{\cal V}}}.
\end{equation}

LA type potential:
\begin{equation}
\xi(-m{\cal V})=\frac{1}{2\pi}\int_{\cal V}^b\frac{P(h){\rm d}h}{\sqrt{h-{\cal V}}}.
\end{equation}

In the LA case we also have a constraint and the displacement can be expressed with
the period:
\begin{equation}
X(v)=\frac{1}{2\pi}\int_0^b{\rm d}h\,\frac{P(h)\sqrt{h}}{h+v^2}.
\label{constraint1}
\end{equation}

\subsection{Examples}
\label{a9}

For MR type potentials we take the example
\begin{equation}
U(x)=\frac{mg^2}{\sinh^2(x/\ell)},
\label{MRex}
\end{equation}
where $g$ is a constant with dimension of velocity and $\ell$ is the unit of length.
The inverse function is
\begin{equation}
\xi(U)=\ell\,{\rm arcsinh}\left(\sqrt{\frac{mg^2}{U}}\right).
\end{equation}
For this example the scattering data can be computed analytically and we find
\begin{equation}
X(v)=\frac{\ell}{2}\ln\left(1+\frac{g^2}{v^2}\right).
\end{equation}

For the LA case we take the example
\begin{equation}
U(x)=-\frac{mg^2}{\cosh^2(x/\ell)},
\label{LAex}
\end{equation}
\begin{equation}
\xi(U)=\ell\,{\rm arccosh}\left(\sqrt{-\frac{mg^2}{U}}\right).
\end{equation}
The scattering data are
\begin{equation}
P(h)=\frac{\ell\pi}{\sqrt{h}},\qquad\qquad
X(v)=\frac{\ell}{2}\ln\left(1+\frac{g^2}{v^2}\right).
\end{equation}
We see that the displacement is exactly the same for the two above cases.

For RC type potentials (see Fig. \ref{fig3}) we take
\begin{equation}
U(x)=mB\,\frac{\xi-{\rm e}^{x/\ell}}{({\rm e}^{x/\ell}-1)^2},
\end{equation}
where $B>0$ is a constant with dimension ${\rm velocity}^2$, $\ell$ is the length unit
and $\xi>1$ is a dimensionless constant.

For small $x$
\begin{equation}
U(x)\approx \frac{m\ell^2 B(\xi-1)}{x^2}
\end{equation}
and for large $x$
\begin{equation}
U(x)\approx -mB{\rm e}^{-x/\ell}.
\end{equation}
The potential vanishes at $x=\ell\ln\xi$ and its minimum is at $x=x^*=\ell\ln(2\xi-1)$:
\begin{equation}
U(x^*)=-mb=-\frac{mB}{4(\xi-1)}.
\end{equation}
The two partial inverse functions are
\begin{equation}
\xi_1(U)=\ell g_1\left(\frac{U}{mB}\right),\qquad\qquad
\xi_2(U)=\ell g_2\left(\frac{U}{mB}\right),
\end{equation}
where
\begin{equation}
g_1(\omega)=\ln\frac{2(\omega-\xi)}{2\omega-1-\sqrt{1+4\omega(\xi-1)}},\qquad\quad
\omega\geq-\frac{1}{4(\xi-1)},
\end{equation}
\begin{equation}
g_2(\omega)=\ln\frac{2(\omega-\xi)}{2\omega-1+\sqrt{1+4\omega(\xi-1)}},\qquad\quad
0\geq\omega\geq-\frac{1}{4(\xi-1)}.
\end{equation}
Again, the scattering data can be calculated analytically:
\begin{equation}
P(h)=\frac{\ell\pi}{\sqrt{h}}-\frac{\ell\pi}{\sqrt{h+B\xi}},
\end{equation}
\begin{equation}
X(v)=\ell\hat X\left(\frac{v}{\sqrt{B}}\right)
\end{equation}
with
\begin{equation}
\begin{split}
\hat X(u)&=\ln\frac{2u^2-1+\sqrt{1+4k}}{2\xi-1+\sqrt{1+4k}}+\ln\frac{1+4k+\sqrt{1+4k}}
{8u^4}\\
&+\frac{u}{\sqrt{u^2-\xi}}\left(\ln\frac{1+\alpha_1\sqrt{u^2-\xi}}
{1-\alpha_1\sqrt{u^2-\xi}}
+\ln\frac{1+\alpha_2\sqrt{u^2-\xi}}{1-\alpha_2\sqrt{u^2-\xi}}\right),
\end{split}
\end{equation}
where
\begin{equation}
\alpha_1=\frac{1}{u(2\xi-1)},\qquad
\alpha_2=\frac{\sqrt{1+4k}-1}{u(\sqrt{1+4k}+2\xi-1)},\qquad k=u^2(\xi-1).
\end{equation}
Note that $\hat X(u)$ is real for all $u>0$, for $u^2<\xi$ we can use the identity
\begin{equation}
\frac{1}{\sqrt{u^2-\xi}}\ln\left(\frac{1+\alpha\sqrt{u^2-\xi}}{1-\alpha\sqrt{u^2-\xi}}\right)=
\frac{2}{\sqrt{\xi-u^2}}\arctan(\alpha\sqrt{\xi-u^2}).
\end{equation}

\subsection{Large distance and low energy asymptotics}
\label{a10}

\subsubsection{MR type potentials}

Let us assume that (as in our examples) the inverse function can be expanded for small $U$
(which corresponds to large $\xi$) as
\begin{equation}
\xi(U)=-{\cal L}\ln\frac{U}{mu_o^2}+\left[\hat\alpha
+\hat\beta\ln\frac{U}{mu_o^2}\right]\frac{U}{mu_o^2}+\dots,
\end{equation}
where ${\cal L}$, $u_o$ and $\hat\alpha,\ \hat\beta$ are constants and
the neglected terms are higher powers of $U$ with coefficients that are polynomials in
$\ln\frac{U}{mu_o^2}$. In this case the low energy expansion of the scattering 
displacement is of the form
\begin{equation}
X(v)={\cal L}\ln\frac{u_o^2}{4v^2}+\frac{2v^2}{u_o^2}\left[\hat\alpha
+\hat\beta\left(\ln\frac{4v^2}{u_o^2}-1\right)\right]+\dots
\end{equation}
plus higher terms in $v^2$ with logarithmic coefficients.
The relation between the two expansions is perturbative (also for the higher terms).
In our $1/\sinh^2$ example
\begin{equation}
{\cal L}=\frac{\ell}{2},\qquad u_o=2g,\qquad \hat\alpha=\ell,\qquad \hat\beta=0.
\end{equation}

\subsubsection{LA type potentials}

Here we assume an expansion of the form
\begin{equation}
\xi(-V)=-{\cal L}\ln\frac{V}{mu_o^2}+{\rm O}(V).
\end{equation}
The corresponding low energy expansion of the scattering data is 
\begin{equation}
P(h)=\frac{2\pi{\cal L}}{\sqrt{h}}+p_o+{\rm O}(\sqrt{h}),\qquad\quad
X(v)={\cal L}\ln\frac{u_o^2}{4v^2}-\frac{vp_o}{2}+{\rm O}(v^2),
\end{equation}
where the constant $p_o$ is non-perturbative and is given by the formula
\begin{equation}
p_o=-\left\{\frac{4{\cal L}}{\sqrt{b}}-2\sqrt{m}\int_0^{mb}\frac{{\rm d}V}{\sqrt{V}}\left[
\xi^\prime(-V)-\frac{{\cal L}}{V}\right]\right\}.
\end{equation}
In our $-1/\cosh^2$ example
\begin{equation}
{\cal L}=\frac{\ell}{2},\qquad u_o=2g,\qquad p_o=0.
\end{equation}

\subsubsection{RC type potentials}

We assume that 
\begin{equation}
\xi_2(-V)=-{\cal L}\ln\frac{V}{mu_o^2}+{\rm O}(V),\qquad\quad
\xi_1(0)={\cal L}z_o.
\end{equation}
The corresponding low energy expansion of the scattering data is 
\begin{equation}
P(h)=\frac{\pi{\cal L}}{\sqrt{h}}+p_o+{\rm O}(\sqrt{h}),\qquad\quad
X(v)={\cal L}\ln\frac{u_o^2}{4v^2}-vp_o+{\rm O}(v^2).
\end{equation}
The non-perturbative constant $p_o$ is given by the formula
\begin{equation}
p_o=-\left\{\frac{2{\cal L}}{\sqrt{b}}+\sqrt{m}\int_0^{mb}\frac{{\rm d}V}{\sqrt{V}}\left[
d^\prime(V)+\frac{{\cal L}}{V}\right]\right\}.
\end{equation}
In our RC example
\begin{equation}
{\cal L}=\ell,\qquad u_o^2=B,\qquad z_o=\ln\xi,\qquad p_o=-\frac{\ell\pi}{\sqrt{B\xi}}.
\end{equation}


  


\vfill\eject

\end{document}